\documentclass[letterpaper,twocolumn,10pt]{article}

\usepackage{usenix,epsfig,endnotes}

\usepackage{algorithm}
\usepackage{algpseudocode}
\usepackage{etoolbox} 

\AtBeginEnvironment{algorithmic}{\useupquotes}

\newcommand{\useupquotes}{%
  \begingroup\lccode`\~=`\'\lowercase{\endgroup\let~}\algoupquote
  \begingroup\lccode`\~=`\"\lowercase{\endgroup\let~}\algoupquotes
  \catcode`\'=\active\catcode`\"=\active
}

\newcommand{\algoupquote}{\mbox{\textquotesingle}}
\newcommand{\algoupquotes}{\mbox{\char`\"}}

\begin{document}

\date{}

\title{Front-Running Protection for Distributed Exchanges using Tamper-Resistant Round Trip Time Measurements}

\author{
{\rm Hannah Atmer}\\
Uppsala University
\and
{\rm Kilian Rausch}\\
Exchange Union\
\and
{\rm Daniel McNally}\\
Exchange Union
}

\maketitle

\thispagestyle{empty}

\subsection*{Abstract}

This paper presents ODIN, a front-running protection system that uses a novel algorithm to measure Round-Trip Time (RTT) to untrusted servers. ODIN is the decentralized equivalent of THOR, an RTT-aware front-running protection system for trading on centralized exchanges. Unlike centralized exchanges, P2P exchanges have potentially malicious peers, which makes reliable direct RTT measurement impossible. In order to prevent tampering by an arbitrarily malicious peer, ODIN performs an indirect RTT measurement that never interacts directly with the target machine. The RTT to a randomized IP address close to the target's IP address in the global routing network is measured to estimate the RTT to the target. We find that ODIN's RTT estimation algorithm provides an accurate, practical, and generic solution for collecting network latency data in a hostile network environment.


\section{Introduction}

ODIN exists to prevent front-running on a decentralized exchange (DEX) conforming to the OpenDEX standard~\cite{opendex}. Front-running is a method used by high-frequency traders to make trades using non-public information. Front-running is an unethical and often illegal~\cite{nasdaq} practice that decreases honest traders' profits. THOR is a system built to protect orders placed on centralized exchanges from front-running. THOR measures the RTT of pings to multiple centralized exchanges and delays sending orders such that the orders arrive at each exchange at the same time~\cite{thor}. THOR prevents high-frequency traders who monitor the exchanges from using information from one exchange to front-run an order on another exchange where the order has not arrived yet due to network latency. Front-running can occur in a DEX when a peer in the network receives information about available trades before other peers receive the information. The simplest way to prevent front-running is to ensure that all peers receive the information simultaneously by introducing an RTT-dependent delay when sending information to multiple peers. However, preventing front-running in a DEX is a more complex problem than preventing front-running on a centralized exchange. The added complexity comes from the potential front-runner's ability to arbitrarily modify the code of the exchange that receives the orders. The front-runner can therefore circumvent all direct measurements of ping latency, e.g. a malicious node operator could install a custom handler for ICMP packets that artificially delays the response to ICMP pings to increase perceived network latency. This would give the malicious node preferential treatment by an algorithm that relies on ICMP pings to measure RTT to ensure that messages reach all peers simultaneously. The same concern applies to all measurements that interact directly with the front-runner's machine or last-hop router; a sufficiently determined adversary can circumvent any direct RTT measurement of a machine that she can arbitrarily modify or influence~\cite{karame1}. ODIN implements a front-running protection algorithm for distributed exchanges that relies on RTT estimates to determine the relative latencies of peers. The exchange node delays sending orders according to the relative latencies of each peer so that all peers receive the order simultaneously. This method ensures that no peer has the opportunity to make trades using information that has not yet arrived at other peers, i.e. to front-run. Given the financial profitability of front-running, we must assume that peers are highly motivated to deceive the front-running protection system. Direct RTT measurement solutions interact directly with the target machine and therefore are vulnerable to tampering. There are no known prior algorithms for measuring RTT in a hostile environment. We outline the design of an RTT estimation algorithm with the following goals:
\newpage
\begin{itemize}
\item Tamper-Resistance: Ensure that the target cannot influence the RTT estimate.
\item Accuracy: Provide high-quality RTT estimates. 
\item Generality: Define the algorithm as a generic RTT estimation algorithm independent of any specific system implementation.
\end{itemize}

We achieve these goals through platform-independent indirect RTT estimation. We generate an accurate estimate of RTT to the target based on the RTT to a machine located near the target in the network. The algorithm is interchangeable with the ping functionality and is easy to implement using the UNIX Traceroute program. Our indirect assessments of the target's RTT generate an RTT estimate that is accurate enough for use in production systems. The RTT estimation algorithm leverages structural properties of internet routing and IP address-block allocation patterns to achieve a close estimate. The crucial insight in this paper is that we can estimate the RTT to a machine with consistent accuracy without interacting with the machine. 

We have implemented ODIN, the front-running protection algorithm that uses indirect RTT estimates, as part of an open source P2P cryptocurrency trading client~\cite{exchangeunion}. However, ODIN and its RTT estimation algorithm are sufficiently general to be implementable in any language for any networked system. Overall, this paper describes the design of a tamper-resistant front-running protection algorithm for DEXs and makes the following contributions:
\begin{itemize}
\item A front-runner on the DEX cannot deceive ODIN's RTT estimation algorithm by modifying the assessed machine or by DDoS of the last-hop router.
\item The RTT estimation algorithm is accurate despite never interacting directly with the target machine, which allows for stealthy measurement of RTT and a way to estimate RTT to machines that are not reachable from the public internet.
\item ODIN showcases the first RTT estimation algorithm designed for distributed systems operating in hostile environments, enabling the development of new kinds of distributed systems.
\end{itemize}

The rest of this paper is organized as follows. Section 2 provides a broad overview of our approach and describes the tamper-resistant components of ODIN in detail. Section 3 describes our implementation of ODIN. Section 4 presents the results of assessing the accuracy of ODIN's RTT estimation algorithm for a set of randomly chosen machines on the internet. Section 5 surveys related systems. Section 6 describes future work, and Section 7 summarizes our contributions.

\section{The ODIN Algorithm}

The central observation that allows for accurate RTT estimation is that the adversary cannot tamper with the assessment without controlling an Autonomous System~\cite{as} or an entire /24 subnet and that neither of these resources is available for use by individuals~\cite{ripe}. This reality of the modern internet makes ODIN's RTT estimation algorithm a robust tool for RTT measurement. The threat model is a high-frequency trader who can co-locate her DEX node's machine with peers in the DEX network or arbitrarily modify her node to feign different latencies at different times to deceive the front-running prevention mechanism. The trader may use packet floods or other means to increase response latency from the last-hop router that serves her machine. The malicious trader is profit-motivated. A trader attempting to front-run should meet an overwhelming time or cost obstacle to making a profit. However, a trader whose node happens to run on a server far away from the majority of other nodes should not be penalized. There may be orders of magnitude differences in naturally occurring latencies between peers on the DEX. For someone running a peer in the same data center as another peer, the RTT could be as low as 0.5ms~\cite{ladis}. For a peer physically located on the opposite side of the earth, the RTT can exceed 300ms. All measurements interacting directly with a potentially malicious node can be deceived by artificially delayed response times. Whether it is by slowing an application's response to ping packets or by modifying the kernel's network driver to respond slowly to new connections, there is always a way to deceive a measurement probe as long as the adversary has root or physical access to the server~\cite{netpaper}\cite{karame2}. In order to prevent an adversary from being able to deceive the assessment algorithm, we measure the RTT to a machine that will have an RTT similar to the target's RTT. We find the IP address of this machine by finding the path to an IP address in the same /24 subnet as the target. This algorithm produces an approximate RTT that is accurate enough for production use.

\subsection{Sequential probing is the only way to discover the path}
When a packet is routed via the internet, the packet's end-to-end path is unknown to the sender. Sequentially probing the path is the only way to discover the IP address of the last-hop router that serves the target's /24 subnet using only public information. The Traceroute~\cite{trace} program discovers this path by sending a series of probes addressed to the target IP address. The first probe packet has the Time-To-Live (TTL) field set to 1 so that the first hop on the path will send a response packet rejecting the probe due to its expired TTL field. The second packet has the TTL field set to 2, and so on, such that each router along the path to the target sends a response packet rejecting the probe due to its expired TTL field (see Figure 1). No packets reach the target, yet the sender receives a series of response packets that describe the route to the target. The protocol most commonly used by Traceroute is ICMP. ICMP packets can be sent without superuser privileges, and most routers will respond to ICMP requests. 
\begin{figure}
\includegraphics[scale=.16]{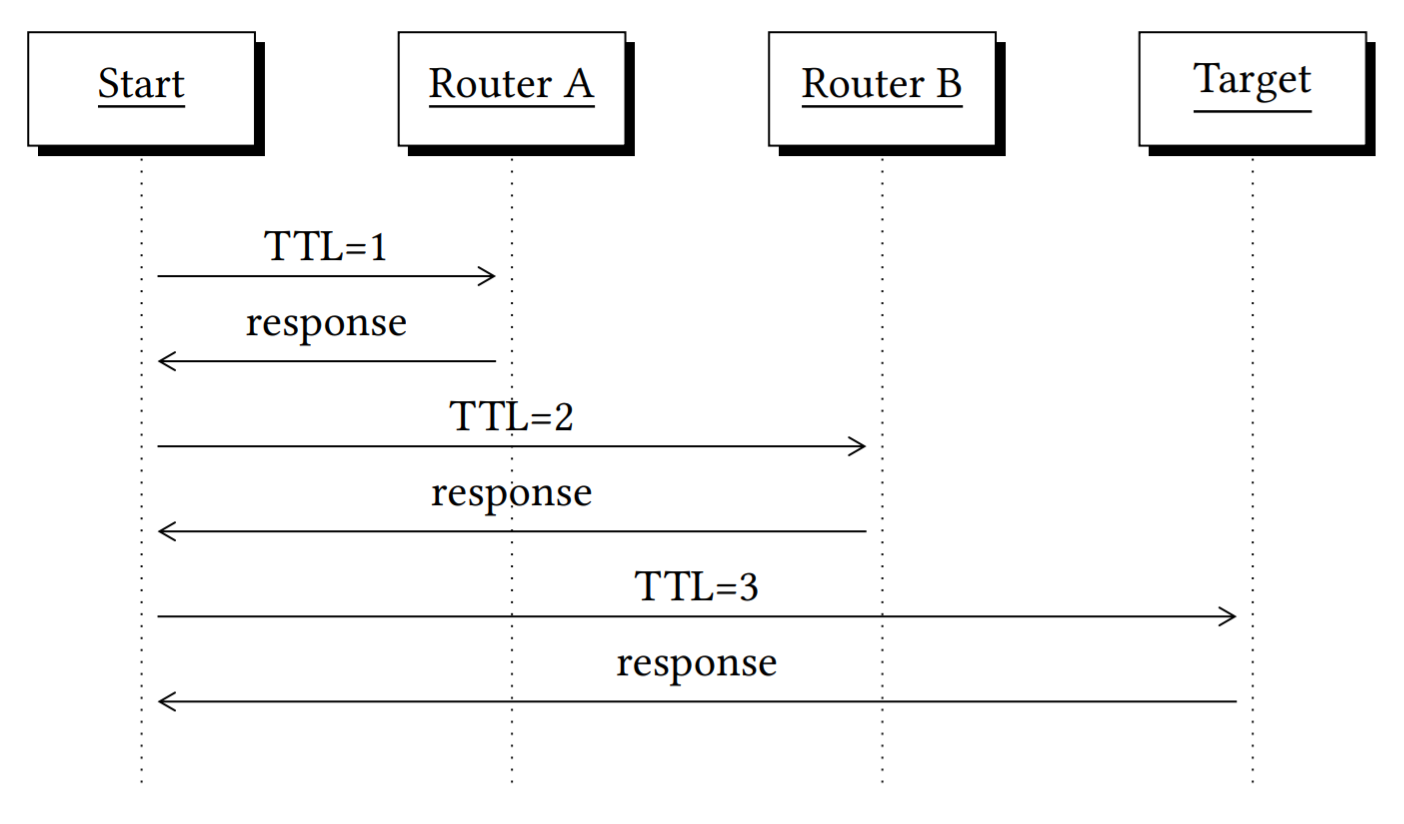}
\caption{Route discovery protocol}
\end{figure}

\subsection{We measure RTT to your neighbor, not to you}

An adversary could deceive the route discovery algorithm by sending forged IP packets claiming to be an intermediary between the penultimate node and the target node. In order to secure ODIN from this deception, we randomize the last octet of the target's IP address and measure latency to this address instead. The RTT to an IP address in the same class C network as the target's IP will usually be very similar to the RTT to the target since a single provider almost always manages class C networks. This reality means that the adversary would need to own an entire class C network, i.e. all 256 addresses in the same /24 subnet, to deceive the measurement consistently. Only large corporations and governments are allowed to own class C networks~\cite{ripe}. The tamper-resistance guarantee of ODIN assumes that the adversary cannot use an entire class C network to deceive the RTT measurement. If the selected neighbor is not reachable, we use the last reachable node on the path since it will likely route to the target, and the final hop rarely adds much to the RTT.

\subsection{Random assessment intervals}
The adversary must not know when the next assessment will take place since the adversary could easily, and at low cost, spawn many short-lived nodes on the same subnet as her DEX instance during the assessment to circumvent the defense described in section 2.2. Alternatively, the adversary could DDoS the router that routes to her /24 subnet at the exact time of the assessment to slow response times. A very short-running DDoS would probably evade detection. The assessment interval is determined with a cryptographically secure pseudorandom number generator to prevent the adversary from predicting the assessment time.

\subsection{Update RTT estimates with incremental increase, immediate decrease}
The ODIN algorithm is resilient against DDoS attacks via strategic adjustment of each peer's RTT estimate. We consider the scenario of an adversary Alice who uses a botnet to DDoS the router that routes to the /24 subnet of her OpenDEX instance, such that the router's performance suffers, i.e. each packet routed by the router takes $\epsilon$ additional seconds. Alice's motivation for this attack is that peers will send orders to her peer faster once they perceive a slower RTT. Alice will then have an opportunity to front-run the orders. Bob's OpenDEX instance, a peer of Alice's OpenDEX instance, will perceive an RTT to Alice's machine that is $\epsilon$ seconds slower than the actual RTT $R$. Alice now stops the DDoS attack on the router in order to allow messages to reach her faster. Bob's OpenDEX instance will perform another RTT assessment in $T$ seconds, but during those $T$ seconds, his OpenDEX instance will behave as if the RTT time to Alice's OpenDEX instance is $R + \epsilon$. Alice gains $T$ seconds, during which Bob's OpenDEX instance sends orders to her OpenDEX instance $\epsilon$ seconds faster. Therefore Alice has $\epsilon$ seconds during which she has the opportunity to front-run Bob's orders and skim his profit. Without preventative measures, Alice can repeat this process every $2T$ seconds such that she has the opportunity to front-run at most $\frac{1}{2}$ of all orders she receives. In order to mitigate this type of attack, we make only tiny upward adjustments of $\delta$ milliseconds to the RTT estimate when a peer's RTT seems to increase. If the RTT assessment shows that the RTT has decreased, we immediately decrease the RTT estimate to the most recently seen RTT. Therefore, Alice must run her DDoS attack for $\frac{\epsilon}{\delta}$ seconds to receive the same preferential treatment from peers, and as soon as she stops the attack she still has only $T$ seconds to take advantage of her deception. If $\delta$ is small enough, Alice will not be able to make a profit from front-running since the cost of performing the DDoS attack will exceed the possible profit from front-running. Choosing a good value for $\delta$ takes into consideration the attacker's potential profit from deceiving the RTT measurement as well as on the inefficiency of forcing the OpenDEX instance's RTT estimates to converge slowly to an accurate estimate, since orders will be sent slower to new peers whose RTT estimate has not yet converged to an accurate value. This feature of the ODIN algorithm is only relevant in contexts involving repeated RTT estimates of the same peer; it can be omitted when the RTT estimation algorithm is used in other contexts.

\section{Implementation}
We have implemented ODIN as a front-running prevention feature~\cite{pr} of an open-source decentralized trading node. The front-running prevention system is easily enabled or disabled to support traders who want optimal profit via front-running protection and traders who prefer optimal trade completion speed. We set an arbitrary maximum delay of 300 milliseconds rather than the RTT to the slowest peer. The actual delay in sending messages is the maximum delay minus the RTT estimate, which limits ODIN's effect on throughput and prevents peers from being able to influence delays for other peers.

\hfill \break
The algorithm uses the following state:
\begin{itemize}
\item rtt\_est: Estimated round trip time. The initial value should be near 0.5ms to prevent the possibility of circumvention via datacenter co-location~\cite{ladis}.
\item delta ($\delta$): The fixed rate of incremental increase of the RTT estimate
\item max\_interval: Upper bound on the amount of time between RTT measurements.
\end{itemize}
\hfill \break
The algorithm for tamper-resistant RTT estimation is as follows:
     \begin{enumerate}   
        \item Replace the last octet of the target IP address with a random value
        \item Measure the RTT to a machine near the target in the network and therefore has a similar RTT.
        \item Randomize the assessment interval so that the time of the next latency assessment is not predictable
    \end{enumerate}
\hfill \break
The algorithm uses the following helper functions:
\begin{verbatim}
/* Runs a traceroute to the IP address.
Returns a list of nodes in the path and the 
RTT to each node in the path. */
string[] traceroute(string addr)

/* Returns a cryptographically secure
pseudorandom number between 0 and max_value */
int crypto_random(int max_value)
\end{verbatim}
\hfill \break
The RTT estimation algorithm pseudocode is as follows:
\begin{verbatim}
float estimateRTT(string addr, float rtt_est) {
  float delta = .1
  string[] octets = addr.split('.')
  octets[3] = crypto_random(255)
  addr = '.'.join(octets)
  trace_lines = traceroute(addr)
  trace_lines.reverse()

  for ip, time in trace_lines {
    if (time) { /* is reachable */
      if (time < rtt_est)
        return time
      else
        return rtt_est + delta
    }
  }
}
\end{verbatim}
\hfill \break
The ODIN algorithm pseudocode is as follows:
\begin{verbatim}
void ODIN(string addr) {
  string addr;
  float rtt_est = 0 /* assume co-located */
  float max_delay = .3
  int max_interval = 180 
  sleep(crypto_random(max_interval))
  rtt_est = estimateRTT(addr, rtt_est)
  /* send order in max_delay-rtt_est seconds */
}
\end{verbatim}

\section{Evaluation}
In this section, we evaluate the accuracy of the ODIN RTT measurement system. We examine ODIN's RTT estimates for randomly generated IP addresses and show that ODIN can accurately and precisely estimate RTT. We evaluate the RTT estimation algorithm in two modes: strict for high accuracy and permissive for improved tamper resistance.

\subsection{Setup}
We analyse the RTT estimation algorithm using 1665 randomly generated reachable IP addresses. We determine whether or not a target IP address is reachable using ping. Upon the success of the ping, we perform two traces: one to the target (A) and one to a random IP address in the same /24 subnet (B). Verifying reachability via ping before performing the traces causes intermediary routers to cache the route, so caching will not bias the difference in the latency data from the two traces. We recorded the RTT to both IP addresses and the RTT to each of their last-hop routers. B does not need to be reachable since the ODIN algorithm can instead use the last reachable node for its estimate. The additional error caused by this strategy is limited since if the last reachable node is not the target IP address, then it is necessarily the node responsible for the target's /24 subnet~\cite{rfc}. The RTT estimate has two possible outcomes: B is reachable, and B is not reachable. Therefore we define two operating modes for the RTT estimation algorithm: strict mode, which requires that B is reachable to make an estimate, and permissive mode, which uses the penultimate node of B for the estimate when B is not reachable. Strict mode will provide more accurate estimates since it will include the latency added by the last hop from the /24 subnet router to a machine at a representative IP address in its subnet.

\subsubsection{Strict Mode}

\begin{figure}
\includegraphics[scale=.35]{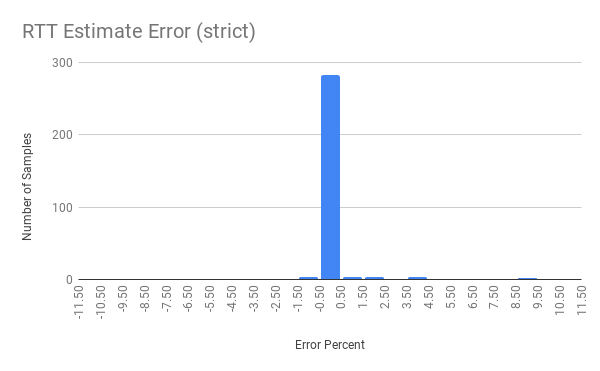}
\caption{Histogram of RTT estimate error when B is reachable. 302 samples.}
\end{figure}

Figure 2 shows the accuracy of the RTT estimation algorithm in strict mode. 94\% of the RTT estimates were within 0.5\% of the actual RTT. This result shows that the RTT estimate is exceptionally accurate when B is reachable. In practice, if A is reachable, then there will usually be other reachable IP addresses in A's subnet. Strict mode estimation is best for ensuring high-accuracy results, and it is trivially implemented by probing until a reachable IP address is found in the target's subnet.


\subsubsection{Permissive Mode}
Figure 3 shows the accuracy of the RTT estimation algorithm in permissive mode. 90\% of the RTT estimates were within 15\% of the actual RTT.

\begin{figure}
\includegraphics[scale=.35]{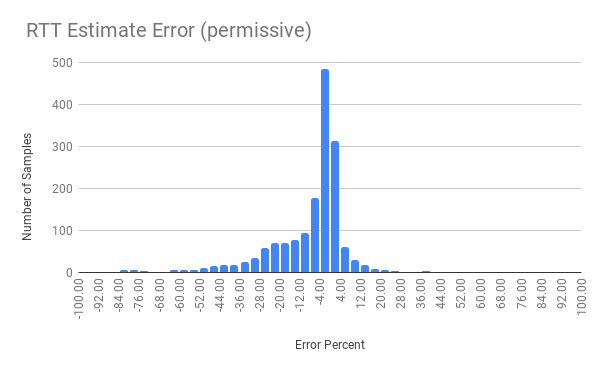}
\caption{Histogram of RTT estimate error when allowing estimates to use the penultimate node on the path to B when B is not reachable. 1665 samples.}
\end{figure}

Given that the permissive mode of the RTT estimation algorithm differs from the strict mode only in that it does not require the last hop latency to be included in the estimate, we infer that the fat tail to the left is due to the missing latency from the last hop. If the type of subnet is known, e.g. residential vs cloud, or prior RTT assessments produced a measure of the RTT difference between any IP address in the target subnet and its router, then a last hop latency estimate can be added to the estimate in order to increase the accuracy. Running ODIN in permissive mode improves assessment speed, stealth, and tamper resistance.

\subsection{Summary} 
In this section, we evaluated the ODIN RTT estimation algorithm's accuracy in both strict and permissive modes. Our evaluation indicates that ODIN achieves highly accurate estimates of RTT in strict mode and reasonably accurate estimates in permissive mode. ODIN's RTT estimation algorithm provides a generic, tamper-resistant, and accurate alternative to pings for determining RTT.

\section{Related Work}
Ping is a program for measuring RTT. Ping uses an ICMP request to elicit a response from a host or router~\cite{ping}. Ping is not tamper-resistant. THOR~\cite{thor} is a trading platform created by Royal Bank of Canada's Capital Markets division to protect their orders from front-running by high-frequency traders. THOR measures pings to centralized exchanges and sends orders with intentional delays. THOR is the inspiration for ODIN, but the ODIN algorithm differs significantly due to the system architecture difference between centralized and decentralized exchanges. Front-running is also a problem for DEXs that make on-chain trades~\cite{flash}\cite{zhou}\cite{sok}. For example, Yunhao et al. built a system for preventing front-running on blockchains by enforcing ordered time slots that prevent byzantine nodes from influencing the order of transactions~\cite{yunhao}. While blockchain-based DEXs are in the same domain as the OpenDEX network, they deal with the problem of front-running on the blockchain itself, i.e. on layer 1. In contrast, OpenDEX orders are gossipped on layer 3, i.e. in a distributed system specialized for trading, which commits trades to the blockchain in batches via the Lightning Network and therefore is concerned only with the network latency between OpenDEX nodes.

\section{Future Work}
Future development of the ODIN algorithm will include automatic tuning of $\delta$ since the optimal value of $\delta$ is dependent on the potential financial profit from front-running.

\section{Conclusion}
This paper has demonstrated that it is possible to estimate RTT to any machine on the internet without direct interaction. We believe the ODIN algorithm is a valuable addition to any distributed systems engineer's toolkit. 

{\footnotesize \bibliographystyle{unsrt}
\bibliography{usenix}}

\end{document}